\documentclass[preprint2]{aastex}

\usepackage{natbib}
\usepackage{amsmath}

\shorttitle{Spherical Plasma Dynamo Experiment}
\shortauthors{Spence et al.}

\begin{document}


\title{A Spherical Plasma Dynamo Experiment}


\author{E. J. Spence} \affil{Department of Physics,
  University of Toronto, 60 St.\ George St., Toronto, Ontario, Canada
  M5S 1A7.}

\author{K. Reuter} \affil{Max-Plank-Institut f\"ur Plasmaphysik,
  EURATOM Association,
  Boltzmannstra{\ss}e 2, D-85748 Garching, Germany.}

\and

\author{C. B. Forest} \affil{Department of Physics, University of
  Wisconsin-Madison, Madison, WI 53706.} \email{cbforest@wisc.edu}

%% are identified by the \altaffilmark after each name.  Specify
%% alternate affiliation information with \altaffiltext, with one
%% command per each affiliation.

%\altaffiltext{1}{present address: Department of Physics, University of
%  Toronto, 60 St.\ George St., Toronto, Ontario, Canada M5S 1A7}

%% Mark off your abstract in the ``abstract'' environment. In the
%% manuscript style, abstract will output a Received/Accepted line
%% after the title and affiliation information. No date will appear
%% since the author does not have this information. The dates will be
%% filled in by the editorial office after submission.

\begin{abstract}
We propose a plasma experiment to be used to investigate fundamental
properties of astrophysical dynamos.  The highly conducting,
fast-flowing plasma will allow experimenters to explore systems with
magnetic Reynolds numbers an order of magnitude larger than those
accessible with liquid-metal experiments.  The plasma is confined
using a ring-cusp strategy and subject to a toroidal differentially
rotating outer boundary condition.  As proof of principle, we present
magnetohydrodynamic simulations of the proposed experiment.  When a
von K\'arm\'an-type boundary condition is specified, and the magnetic
Reynolds number is large enough, dynamo action is observed.  At
different values of the magnetic Prandtl and Reynolds numbers the
simulations demonstrate either laminar or turbulent dynamo action.
\end{abstract}

%% Keywords should appear after the \end{abstract} command. The
%% uncommented example has been keyed in ApJ style. See the
%% instructions to authors for the journal to which you are submitting
%% your paper to determine what keyword punctuation is appropriate.

\keywords{MHD, plasmas, stars: magnetic fields, ISM: magnetic fields}

\section{Introduction}

One of the central problems of astrophysical fluid dynamics is the
magnetohydrodynamic (MHD) dynamo, the process by which the motion of
an electrically conducting fluid amplifies a small seed magnetic field
until the field becomes dynamically important.  Astrophysical dynamos
can be categorized into two types: the large-scale dynamo, where the
scale of the generated magnetic field is the same as or larger than
the system generating it, such as in planets or stars, or the
small-scale dynamo, where the scale of the magnetic field is much
smaller than the system size, such as in warm interstellar
medium~\citep{Kulsrud.ARAA.1999}.  Of particular physical importance
in this context is the system's magnetic Prandtl number, $Pm =
\nu/\eta$, where $\nu$ is the kinematic viscosity of the fluid and
$\eta$ its magnetic diffusivity, whose value indicates the scale of
the velocity field at the resistive-diffusive cutoff.  When $Pm \ll 1$
the magnetic-diffusion scale is in the hydrodynamic inertial range and
the primary interactions are at that scale.  If a dynamo occurs in
this situation it is often large scale, though recent evidence
indicates that a small-scale dynamo may also be
possible~\citep{Iskakov.PRL.2007, Schekochihin.NJP.2007}.  Conversely,
$Pm \gg 1$ gives a magnetic-diffusion scale below the
viscous-diffusion cutoff, permitting the stretching and folding of the
magnetic field at small scales and allowing the opportunity for a
small-scale dynamo.

The small-scale dynamo has only been studied theoretically and through
numerical simulations (see \citet{Haugen.PRE.2004,
  Schekochihin.NJP.2007} and references therein).  The simulations
tend to be done in periodic boxes with random non-helical forcing, in
the absence of a mean flow, allowing the study of the dynamo under
homogeneous and isotropic fluctuating conditions.  Most simulations
have been done in the range of $Pm \sim 1$, meaning with identical
viscous and magnetic dissipation scales, though there have been
notable exceptions where much higher values of $Pm$
\citep{Schekochihin.APJ.2004, Schekochihin.APJ.2002, Kinney.APJ.2000}
and values with $Pm < 1$~\citep{Iskakov.PRL.2007,
  Schekochihin.NJP.2007} have been used.  Despite the large body of
numerical and theoretical work on such dynamos, small-scale dynamos
have never been experimentally studied due to the dearth of fluids for
which $Pm \gg 1$ or $Pm \sim 1$.  The only fluids which satisfy these
criteria are plasmas, and plasma dynamo experiments, to study large or
small-scale dynamos, have not thus far been constructed.

In contrast, several groups have recently been successful in achieving
large-scale dynamo action in experiments using liquid sodium, a fluid
for which $Pm \ll 1$~\citep{Gailitis.PRL.2000, Stieglitz.PF.2001,
  Monchaux.PRL.2007}.  In such experiments the magnetic Reynolds
number, $Rm = v_0 a / \eta$, where $v_0$ is a characteristic speed,
and $a$ a length scale, characterizes the ratio of magnetic field
advection to diffusion.  For idealized laminar velocity fields the
critical value of $Rm$ for magnetic self-excitation, $Rm_{\rm crit}$,
is predicted to be around 100 \citep{Forest.MHD.2002,
  Ravelet.PF.2005}.  To achieve this using liquid sodium in an
experiment of radius $a=0.5$ m requires a mechanical input power of
$P_{\rm mech}\sim 100$ kW.  These flows are very turbulent, however,
and turbulent flows have the unfavorable scaling $Rm\propto (P_{\rm
  mech} a)^{1/3}$.  It would be interesting to study dynamo physics at
a higher order of magnitude, $Rm=1000$ for example, but to achieve
such a value of $Rm$ in a sodium experiment of similar size would
require a mechanical input power of 100 MW!  This is a serious
limitation to addressing higher-$Rm$ dynamo regimes using liquid
sodium.

A different class of fluids which could be used to study dynamo
physics is plasma.  To match the conductivity of sodium a singly
ionized plasma requires an electron temperature of 630 eV, a plasma
temperature only found in fusion experiments.  However, plasma flows
can be efficiently driven to much higher speeds than liquid metals.
Thus for a plasma experiment of similar size ($a = 1$ m) to achieve a
value of $Rm=1000$ would simultaneously require an electron
temperature of only $T_{\rm e}$ = 10 eV and a velocity $v_0$ = 20 km
s$^{-1}$.  These are modest values which can easily be achieved in
many plasma confinement configurations.  Plasma experimenters also
have the novel ability to change the viscosity of their plasmas by
several orders of magnitude.  Such flexibility could allow the
exploration of laminar and turbulent dynamos in a wide range of $Pm$,
from $Pm \ll 1$ to $Pm \gg 1$.  Such variability would be a striking
improvement over liquid-metal experiments.

However, plasmas suffer from an important disadvantage when compared to
liquid metals: they can be difficult to control and confine without
the presence of a magnetic field.  This is a problem for the study of
dynamos, of course, because the presence of a dynamically significant
magnetic field changes the fundamental behavior of the problem under
study.  Those challenges common to most plasma experiments--the need
for thermal confinement to keep the plasma away from container walls
and hot enough to be a good conductor, and some scheme to drive the
flow--must be overcome, in a dynamo experiment, without magnetizing
the plasma.  A carefully designed confinement scheme is needed to meet
such criteria.

In this paper we introduce a multipurpose plasma experiment proposed
by one of the authors~\citep{Forest.APS.2008}.  The experiment is
spherical and based on an axisymmetric ring-cusp confinement strategy.
The confining magnetic field is localized to the periphery of the
experiment and a large, unmagnetized plasma volume is created in the
experiment's core.  The only considered means of injecting energy into
the plasma, once formed, is by forcing the velocity field in the
toroidal direction at the plasma's outer edge.  Unlike spherical
Couette flow experiments, whose outer boundary rotates as a rigid
rotor, the proposed experiment's outer boundary is capable of
differential forcing as a function of angle, meaning, approximately, a
velocity field subject to the boundary condition
$\mathbf{v}(r=a,\theta,\phi) = v_\phi(\theta)\hat{\boldsymbol{\phi}}$,
where $a$ is the radius of the sphere.  The only physical restriction
on the boundary condition is that it remain zero at the poles, meaning
$v_\phi(0) = v_\phi(\pi) = 0$.  By adjusting the plasma such that $Pm
= 1.0$, and choosing $Rm$ appropriately, the experiment should be able
to generate a laminar dynamo, conditions that have never before been
created in the laboratory.  It should also be able to produce
turbulent dynamos by operating high-fluid-Reynolds-number flows with
high values of $Rm$.  It would also not be restricted to dynamo
physics, as a vast set of velocity fields could be generated in a
simply-connected volume, allowing the experimental study of numerous
astrophysical and geophysical phenomena.  Further details of the
proposed experiment are presented in Section~\ref{section:experiment}.

We also present numerical simulations as proof-of-concept of the
proposed experiment, simulating a few of the many possible
configurations using a single-fluid approximation of the plasma.  The
simulations are performed using a parallelized version of a
three-dimensional incompressible non-linear MHD code developed to
simulate the Madison Dynamo Experiment \citep{Bayliss.PRE.2007,
  Reuter.CPC.2008}.  By varying the outer toroidal velocity field
boundary condition different flow regimes have been studied.  In
Section~\ref{section:dynamo} we present a boundary condition which
results in flows which display laminar or turbulent dynamo action
depending on the simulation's value of the magnetic Reynolds and
Prandtl numbers.  We conclude with a discussion of possible avenues of
research for the experiment.

\section{Experimental description}
\label{section:experiment}

\begin{figure}
\plotone{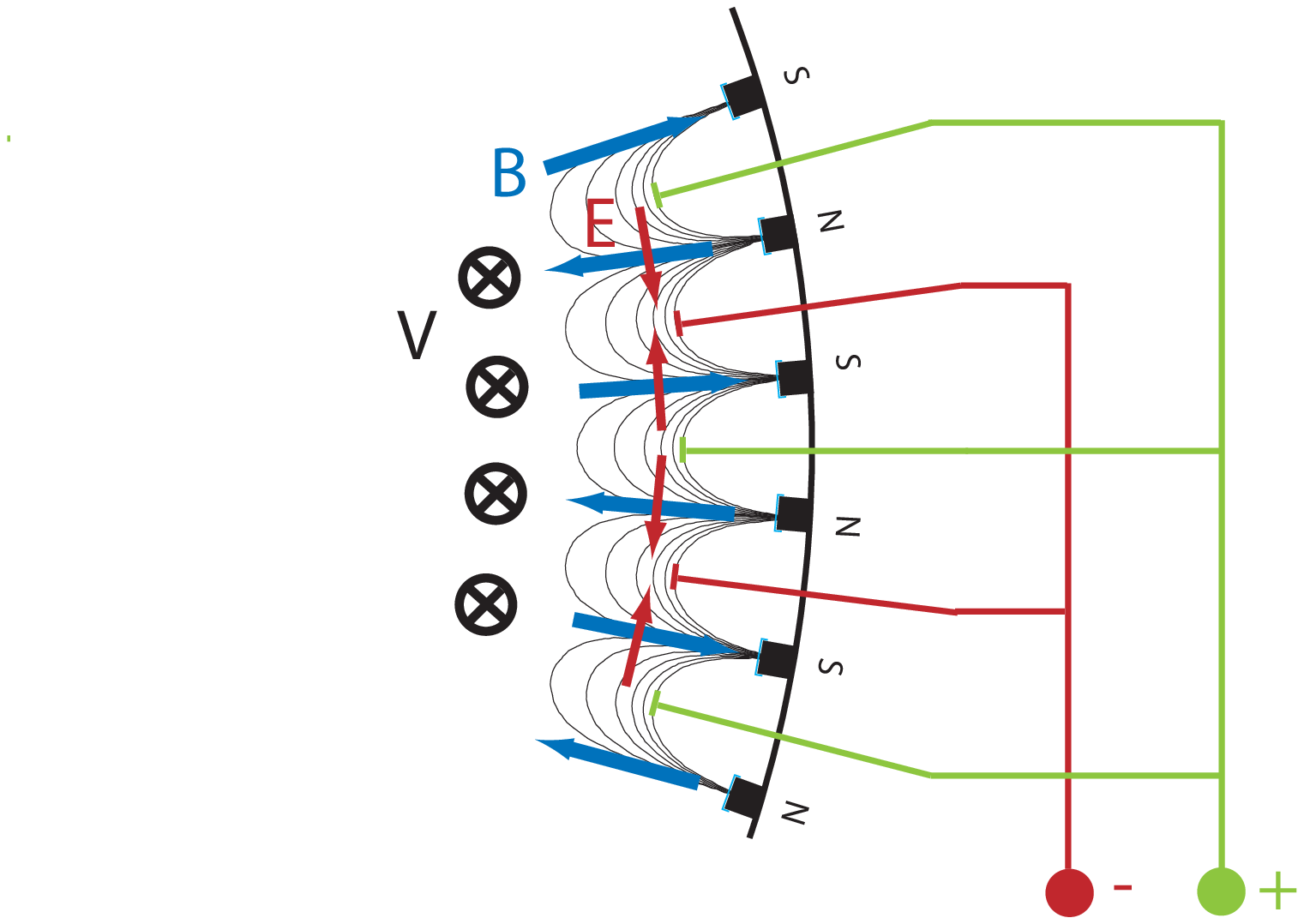}
\caption{Partial vertical cross section of the proposed experiment.
  Rings of permanent magnets, of alternating polarity, line the inside
  of the sphere with their poles oriented radially.  Ring anodes and
  cathodes lie between the magnets.  The resulting
  $\mathbf{E}\boldsymbol{\times}\mathbf{B}$ drift is in the toroidal
  direction.  By varying the potential between the anodes and cathodes
  the forcing at the outer boundary can be customized.  The apparatus
  is axisymmetric.}
\label{fig:fig1}
\end{figure}

The plasma is confined in a spherical geometry using an axisymmetric
ring-cusp strategy, consisting of rings of permanent magnets mounted
to the inner wall of the sphere.  The poles of the magnets are
oriented radially, and the rings have alternating polarity.  A partial
cross section of this configuration is presented in
Figure~\ref{fig:fig1}.  Such a confinement strategy has been used
previously in a cylindrical geometry \citep{Limpaecher.RSI.1973,
  Leung.PF.1976, Lang.JAP.1978, Cho.JVSTA.1988}, and is currently used
in ion sources for neutral beam heating~\citep{Ehlers.RSI.1979} and in
the context of plasma processing~\citep{Pelletier.RSI.1984}.  The
magnetic field is localized to the outer edge of the sphere and drops
to a negligible value within a radial distance on the order of the
space between the magnets.  The experimental volume is essentially
magnetic field free, resulting in a very high-$\beta$ plasma, {\it
  i.e.} $\rho v^2 \gg B^2/\mu_0$, where $\rho$ is the plasma mass
density and $\mu_0$ the magnetic permeability of a vacuum.

Confinement in such a magnetic geometry has been well studied.
Particle confinement is determined by ion-acoustic flows into the
cusps, which can be modeled as a loss area equal to the linear
dimension of the cusp times the ion
gyroradius~\citep{Hershkowitz.PRL.1975}.  Electrons are well confined
by the cusp fields~\citep{Leung.PF.1976}; their primary energy losses
are through the cusp and through ion collisions.  Ion energy
confinement is mostly determined by charge exchange losses resulting
from the relatively high fraction of neutrals in such devices.  The
ionization fraction of previous and current multidipole confinement
systems varies considerably, but is typically $<$ 0.2.

Improved technology could allow the proposed experiment to exceed the
confinement limits of earlier apparatuses.  Modern NeFeB permanent
magnets can generate cusp fields twice as strong as those in previous
cylindrical experiments.  This would result in lower loss rates, due
to the narrowing of the cusps, and increased electron temperature,
which has been shown to scale with cusp magnetic field
strength~\citep{Ehlers.RSI.1982}.  Heating of the plasma could be done
using large area Lanthanum Hexaboride (LaB$_6$) cathodes.  Such
cathodes have previously been used in ion
sources~\citep{Ehlers.RSI.1979, Leung.RSI.1984, Pincosy.RSI.1985}, and
in a variety of plasma experiments \citep{Ono.PRL.1987,
  Darrow.PFB.1990}.  These sources produce significantly more power
than tungsten filaments, which would lead to higher temperatures and
ionization fractions than previous devices.

To inject momentum into the experiment, ring anodes and cathodes are
alternately placed between the ring magnets (Figure~\ref{fig:fig1}).
These generate an electric field that causes an $\mathbf{E}
\boldsymbol{\times} \mathbf{B}$ drift in the toroidal direction.  The
magnitude and direction of this drift can be varied as a function of
angle by changing the electrical potential between the anodes and
cathodes, allowing control over the fluid's outer toroidal velocity
field boundary condition.  Viscous coupling between the magnetized and
unmagnetized regions of the experiment is assumed.

A broad set of experimental regimes can be generated using such a
device.  Some of the specifications of the proposed experiment are
listed in Table~\ref{table:parameters}.  In the context of this study,
we are particularly interested in the magnetic Reynolds and Prandtl
numbers.  The magnetic Reynolds number for a plasma is given by
\begin{equation}
Rm=1.5 \frac{T_{\rm e}^{3/2} U a}{Z},
\end{equation}
where the electron temperature, $T_{\rm e}$, is measured in electron
volts, $U$ is the peak speed of the plasma in km s$^{-1}$, the length
scale $a$ is measured in meters, and $Z$ is the charge of the ions.
High electron temperatures, and the high speeds of plasmas, can lead
to very large values of magnetic Reynolds number.  In an unmagnetized
plasma the viscosity is given by $\nu \sim V_{T{\rm i}}^2 \tau_{\rm
  ei}$, where $V_{T{\rm i}} = 2\sqrt{T_{\rm i}}{m_{\rm i}}$ is the
thermal velocity of the ions, $\tau_{\rm ei}$ is the electron-ion
collision time, $T_{\rm i}$ is the ion temperature and $m_{\rm i}$ is
the ion mass.  The unmagnetized magnetic Prandtl number is then
\begin{equation}
  Pm=0.18 \frac{T_{\rm e}^{3/2} T_{\rm i}^{5/2}}{\mu^2 n},
\end{equation}
where the ion temperature is measured in electron volts, $\mu$ is the
atomic mass number of the ions, and $n$ is the number density in units
of $10^{18}$ m$^{-3}$. The strong dependence on density and ion mass
allows a very large range of magnetic Prandtl number, and by extension
fluid Reynolds number, $Re=Rm/Pm$, allowing the experimenter the
ability to specify whether a given plasma will be laminar or
turbulent.
\begin{deluxetable}{llrl} 
\tablecolumns{4} 
\tablewidth{0pc} 
\tablecaption{Parameters of the Experiment} 
\tablehead{ 
\colhead{Quantity} & \colhead{Symbol} & \multicolumn{1}{r}{Value} & 
\multicolumn{1}{l}{Unit}}
\startdata 
plasma radius & $a$ & 1.5 & m\\
number density & $n$ &  10$^{17}$--10$^{19}$ &  m$^{-3}$\\
ion temperature & $T_{\rm i}$ &  0.5--4 & eV \\
electron temperature & $T_{\rm e}$ & 2--10 & eV \\
peak speed & $U_{\rm max}$  & 0--20  & km s$^{-1}$ \\
ion species & H, He, Ar & 1, 4, 40 & amu \\ 
pulse length & $\tau_{\rm pulse}$ & 5 & s \\
plasma beta & $\beta$ & 10$^4$\\\hline
resistive time & $\tau_\eta$ & 50 & ms \\
magnetic Reynolds number & $Rm_{\rm max}$ & $\sim$ 1000--2000 &  \\ 
Reynolds number & $Re$ & 2.4$\times$10$^{1}$--3.8$\times$10$^{6}$ \\
magnetic Prandtl number  & $Pm$ & 3.0$\times$10$^{-4}$--5.6$\times$10$^{1}$ \\
\enddata 
\label{table:parameters}
\end{deluxetable}

Varying density and ion mass to control the viscosity of the plasma is
not without its tradeoffs, the primary one being the saturation level
of the magnetic field generated by the dynamo, since it is dependant
upon the kinetic energy of the fluid.  Assuming that saturation occurs
when the magnetic energy is in equipartition with the kinetic energy,
$B^2/\mu_0 = \rho U^2$, implies
\begin{equation}
  B_{\rm equipartion} \approx 0.4 \sqrt{\mu n} U\: \mbox{Gauss}.
\end{equation}
At $T_e=10$ eV, $U=$ 20 km s$^{-1}$, $B _{\rm equipartion}$ could be
as high as 240 Gauss with Ar at $n=10^{19}$ m$^{-3}$.  However, as we
will show in Section~\ref{section:dynamo}, simulations indicate that
in saturation the magnetic field magnitude is not this high, but at
best $B^2/\mu_0 \simeq 0.1\rho U^2$.

The lack of magnetic field in the volume of the experiment and a
method to control the velocity field means that this device should
satisfy the criteria needed to create magnetically self-exciting
plasmas.  Its high value of $Rm$ and variable magnetic Prandtl number
would allow it to be used to study physical regimes inaccessible to
liquid-metal experiments.  It would be the first plasma experiment
used to study dynamos, though not the first to be proposed
\citep{Wang.PP.2002}, and its wide range of parameters
(Table~\ref{table:parameters}) should allow it to be useful for the
study of a number of other astrophysical phenomena: magnetorotational
instability, high-$\beta$ instabilities and rotating convection, to
name just a few.  In the next section we present numerical simulations
which demonstrate that this concept should succeed as a dynamo
experiment.  The other topics above will be examined in future
work.

\section{Numerical simulations}
\label{section:dynamo}

A comprehensive simulation of such a plasma experiment would consider
both the electron and ion species, as well as include a careful
treatment of the $\mathbf{E}\boldsymbol{\times}\mathbf{B}$ forcing at
the outer boundary.  As a first approximation, however, we treat the
plasma as an incompressible single fluid.  This is a justified
approximation since the mean free path of the particles is much
shorter than the system size, and sound waves are not relevant.  The
simulation solves the vorticity evolution and magnetic induction
equations using a standard pseudospectral method based on spherical
harmonics, truncated at maximum spherical harmonic degrees $\ell_{\rm
  max}$ and $m_{\rm max}$; details have been given previously
\citep{Bayliss.PRE.2007, Reuter.CPC.2008}.  The
$\mathbf{E}\boldsymbol{\times}\mathbf{B}$ forcing is approximated as a
toroidal velocity field non-zero no-slip boundary condition.  An
electrically-insulating outer boundary is assumed.

For the sake of brevity, here we restrict our attention to two values
of magnetic Reynolds number, $Rm=300, 2000$, where the magnetic
Reynolds number is based on the peak speed of the toroidal outer
boundary condition, which is unity in all simulations.  We examine
several values of $Pm$ to demonstrate that this experiment should be
able to generate both laminar and turbulent dynamos.  Future studies
will examine other regimes of interest.

\subsection{Toroidal boundary conditions}

\begin{figure}
\plottwo{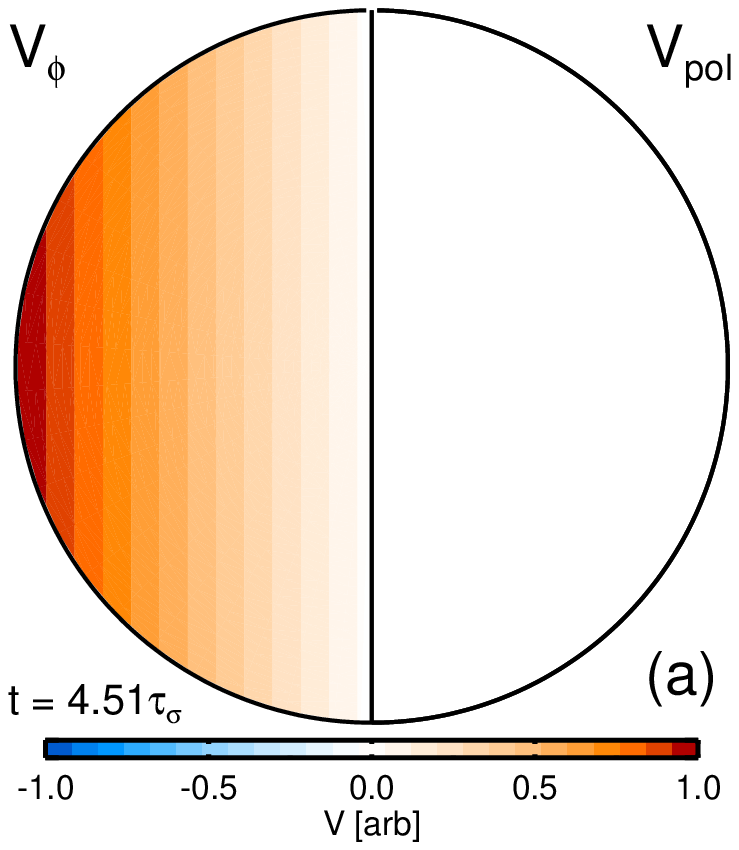}{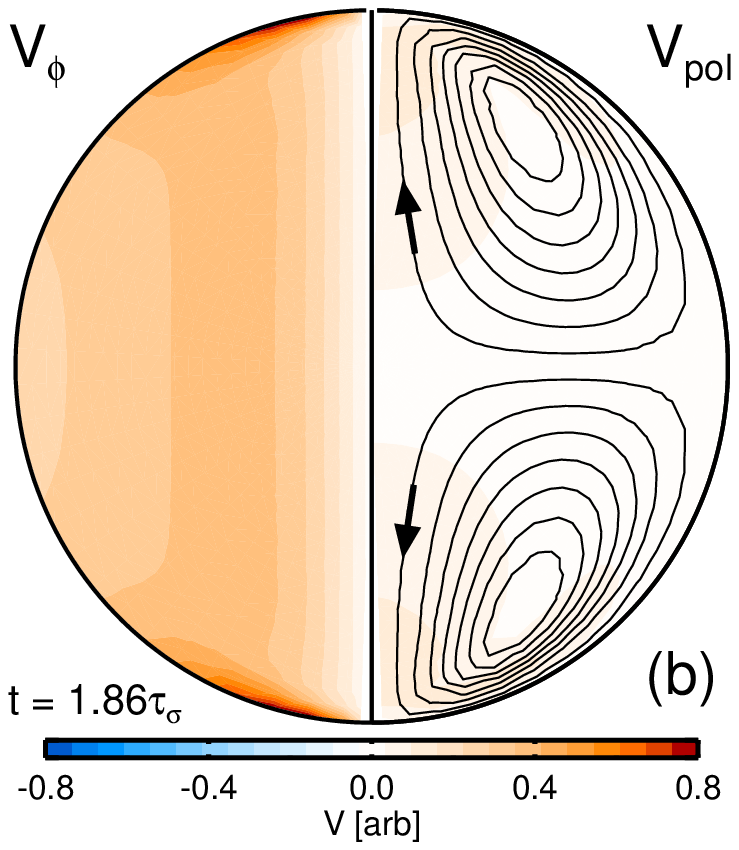}\\ 
\plottwo{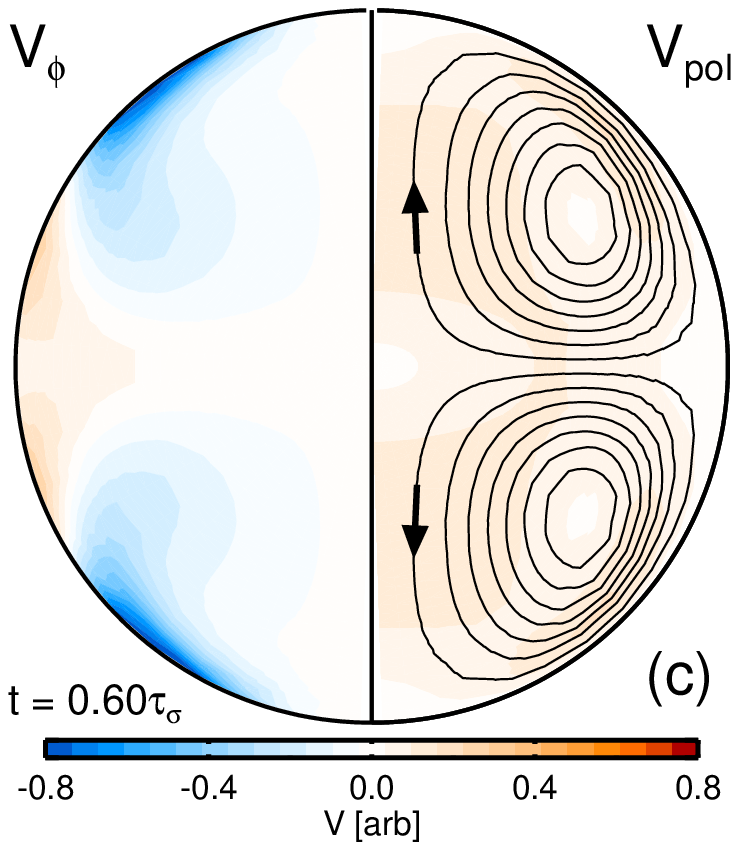}{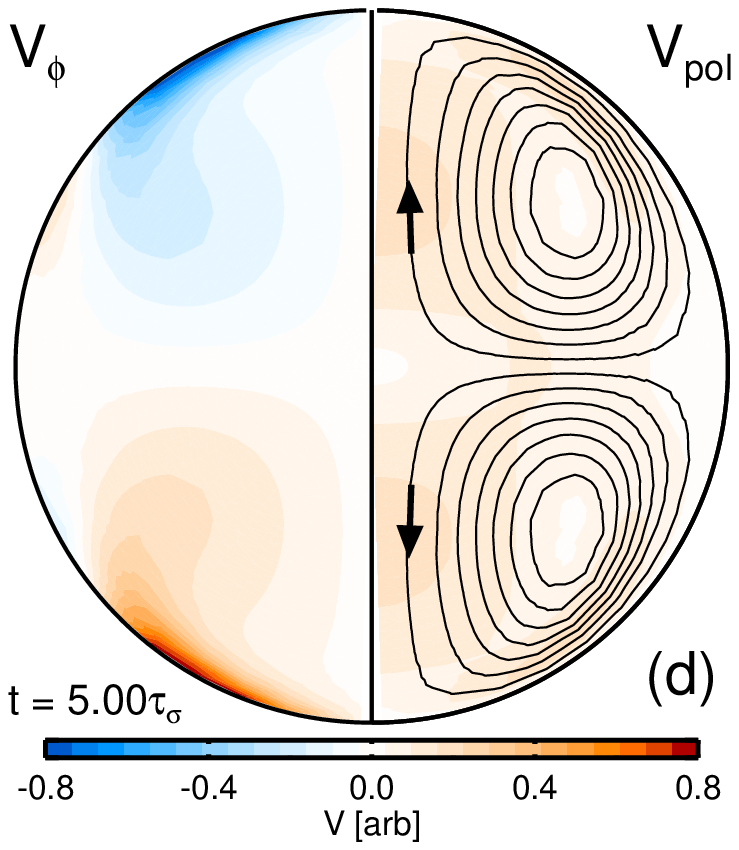}
\caption{Steady-state axisymmetric velocity fields generated by
  different toroidal boundary conditions, with $Rm=300$ and $Pm=1.0$.
  The axis of symmetry runs vertically.  In the left hemisphere are
  the contours of toroidal speed, and in the right hemisphere are the
  contours of the poloidal stream function.  The boundary conditions
  are (a) solid-body rotation, which generates no poloidal flow (b)
  $v_\phi \sim s^{-1}$, which results in a Keplerian profile (c)
  equatorially symmetric, which results in dynamo action and (d) von
  K\'arm\'an (given in Figure~\ref{fig:fig3}), which also generates a
  dynamo.  Note that, as indicated in Figure~\ref{fig:fig3}, the peak
  speed is set to 1.0, but the scale range in figures (b)-(d) has been
  reduced for clarity.}
\label{fig:fig2}
\end{figure}

To demonstrate the flexibility of such an apparatus, four different
axisymmetric steady-state velocity fields, generated using different
outer boundary conditions with $Rm=300$ and $Pm=1.0$, are presented in
Figure~\ref{fig:fig2}.  The first, perhaps the simplest that can be
generated in such a device, is a rigid-rotor-type azimuthal field.
Such a velocity field can be used, in combination with the injection
of light ions into a heavy plasma, to study rotating convection in
plasmas.  The second velocity field, Figure~\ref{fig:fig2}(b), is
generated by a boundary condition that follows a $v_\phi\sim s^{-1}$
dependence for much of its angular extent, where $s$ is the
cylindrical radial coordinate.  The toroidal flow generated by this
boundary does not have a $s^{-1}$ dependence, but rather is Keplerian
for much of the volume of the sphere, with $v_\phi\sim s^{-1/2}$ at
the equator, and as such should be unstable to the magnetorotational
instability (MRI)~\citep{Balbus.AJ.1991A}.  Not surprisingly, when
this flow is exposed to an externally-applied magnetic field an
MRI-like instability appears to develop.

\begin{figure}
\plotone{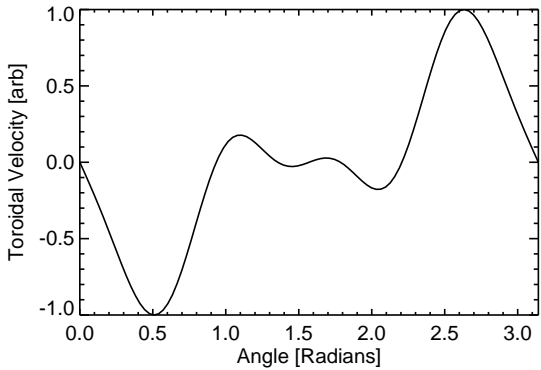}
\caption{Toroidal boundary condition used in the simulations to
  generate a von K\'arm\'an-type flow.  This boundary condition is
  constructed from the boundary values specified in
  Appendix~\ref{app:boundary_values}.}
\label{fig:fig3}
\end{figure}

The velocity fields presented in Figures~\ref{fig:fig2}(c)
and~\ref{fig:fig2}(d) are both magnetically unstable, displaying
dynamo action under the conditions presented.  The first boundary
condition is equatorially symmetric, resulting in a flow with net
kinetic helicity, while the second boundary condition is equatorially
anti-symmetric.  For the remainder of this paper we will focus on the
second of these two cases, which is based on the von K\'arm\'an
flow~\citep{vonKarman.ZAMM.1921}, a flow which has been extensively
studied in the cylindrical geometry \citep{Odier.PRE.1998,
  Bourgoin.PF.2002, Bourgoin.MHD.2004}.  In this case the outer
boundary rotates in opposite directions near the poles of the sphere
and relatively little near the equator, as seen in
Figure~\ref{fig:fig3}.  The boundary condition is constructed from
non-zero outer boundary values for the even-numbered axisymmetric
spherical harmonic components $\ell=2,4,6,8$; these values, and all
the boundary values used in this study, are given in
Appendix~\ref{app:boundary_values}.  This boundary condition results
in laminar and turbulent dynamo action depending on the value of $Rm$
and $Re$ (for this boundary condition the transition to turbulence
begins at $Re \sim 500$, depending on the level of forcing).  The
parameter values used in the simulations presented in this study, and
exemplary physical values of corresponding plasmas, can be found in
Table~\ref{table:run_parameters}.
\begin{deluxetable}{cccccccccccc} 
\tablecolumns{11} 
\tablewidth{0pc} 
\tablecaption{Simulation Parameters \& Plasma Values} 
\tablehead{ 
 & & & \colhead{$T_{\rm e}$} & \colhead{$T_{\rm i}$} & \colhead{$U$} & 
\colhead{$n$} & & \colhead{$B_{\rm max}$} & & & \\
\colhead{$Rm$} & \colhead{$Pm$} & \colhead{$Re$} & \colhead{(eV)} &
\colhead{(eV)} & \colhead{(km s$^{-1}$)} & \colhead{(10$^{18}$ m$^{-3}$)} & 
\colhead{Ion} & \colhead{(G)} & \colhead{$N_r$} & 
\colhead{$\ell_{\rm max}$} & \colhead{$m_{\rm max}$}}
\startdata 
300  & 2.0 & 150  & 4.3 & 3.1 & 20.0 & 13.8 & H  & 3.2 & 200 & 14 & 14  \\ 
300  & 1.0 & 300  & 4.3 & 3.1 & 20.0 & 27.6 & H  & 4.2 & 200 & 14 & 14 \\ 
300  & 0.5 & 600  & 10.0 & 4.0 & 5.7 & 22.7 & He &     & 300 & 18 & 14 \\
2000 & 1.0 & 2000 & 15.9 & 4.0 & 19.1 & 23.0 & He & 12.0 & 600 & 50 & 30
\\ \enddata 
\tablecomments{The radius of the unmagnetized plasma volume, $a$, is
  taken to be 1.1 m.  $N_r$ is the number of radial grid points,
  equally spaced in the range $0 < r \le 1$.  $B_{\rm max}$ refers to
  the maximum saturated magnetic field strength.}
\label{table:run_parameters}
\end{deluxetable} 

\subsection{Laminar dynamo simulations}
\label{section:large_scale}

Among the many appealing features of the proposed experiment is its
ability to create laminar MHD flows at very low values of fluid
Reynolds number, something that is not possible with liquid-metal
experiments.  Though such flows are an imperfect analogy of
naturally-occurring dynamos, since all natural flows are turbulent,
their study can be used to test basic MHD under idealized conditions
as well as the numerical codes which simulate them.
Low-Reynolds-number flows can also be used to study the
laminar-to-turbulent transition, and should allow the observation of
the spontaneous relaminarization of a turbulent flow by a self-excited
magnetic field~\citep{Bayliss.PRE.2007}.  Here we restrict our
examination to a set of laminar dynamos which could be generated in
the proposed experiment.

It might come as a bit of a surprise that a laminar velocity field
generated by a toroidal differentially-rotating outer boundary can
generate a dynamo, since the poloidal component of the velocity field
only results from Ekman circulation, making it difficult to achieve
the balance of toroidal-to-poloidal flow needed for laminar dynamo
action.  Also, since the toroidal flow peaks at the edge of the
sphere, the poloidal field is unable to easily stretch the magnetic
field around the toroidal flow's peak, and thus the stretch-twist-fold
mechanism \citep{Childress} that sustains laminar dynamos is much less
efficient than that of other flows \citep{Dudley.PRSLA.1989}.  This
manifests itself in the comparably large critical magnetic Reynolds
number required for the toroidally-driven flows to self-excite.

\begin{figure}
\plotone{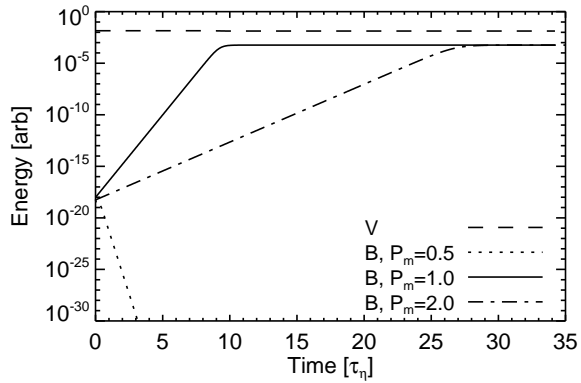}
\caption{Kinetic and magnetic energies versus time for the system
  forced by the boundary condition given in Figure~\ref{fig:fig3},
  with $Rm = 300$ and $Pm = 0.5, 1.0, 2.0$.  The kinetic energy for
  the three simulations is essentially the same.  The $Pm=0.5$
  simulation does not display dynamo action.}
\label{fig:fig4}
\end{figure}

\begin{figure}
\plotone{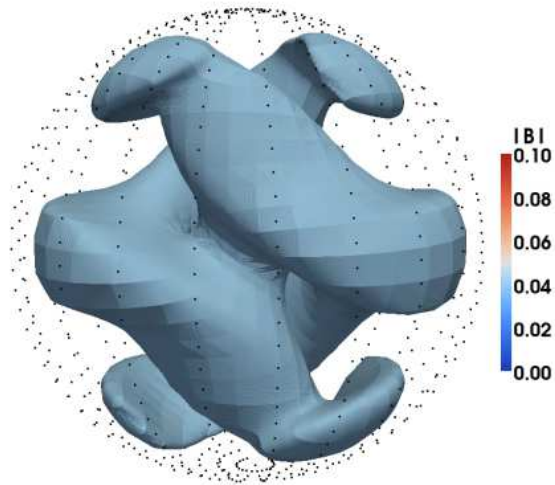}
\caption{Contour of constant magnetic field magnitude during
  saturation, for the $Rm=300$, $Pm=1.0$ laminar dynamo.  The axis of
  symmetry runs vertically.}
\label{fig:fig5}
\end{figure}

The velocity field which results from the von K\'arm\'an boundary
condition, with $Rm=300$ and $Pm=1.0$ (Figure~\ref{fig:fig2}(d)), is
axisymmetric, counter-rotating in the toroidal direction, and has a
poloidal flow which rolls inward at the equator and outward at the
poles.  As such the velocity field is qualitatively similar to the
$s2t2$ flow of \citet{Dudley.PRSLA.1989}, with the important
distinction that the toroidal flow peaks at the sphere boundary.  The
magnetic and kinetic energies of this simulation, as a function of
time in resistive units ($\tau_\eta = a^2 / \eta$), are given in
Figure~\ref{fig:fig4}.  The growth rate of the magnetic energy is very
small, symptomatic of the flow's inefficiency as a dynamo.  The
critical magnetic Reynolds number for this flow, based on a linear
stability analysis, is $Rm_{\rm crit} \simeq 237$.  As is required for
axisymmetric velocity fields by Cowling's theorem
\citep{Cowling.MNRAS.1933}, the excited magnetic field is
non-axisymmetric, dominated by $m=1$ modes; this is a large-scale
dynamo.  Overall the final magnetic field is weak, with the magnetic
energy peaking an order of magnitude below the kinetic energy,
resulting in minimal modification of the velocity field during
saturation.  The large-scale structure of the saturated magnetic field
is given in Figure~\ref{fig:fig5}, demonstrating a geometry similar to
other $s2t2$-type dynamo magnetic fields~\citep{Bayliss.PRE.2007,
  Gissinger.PRL.2008, Gissinger.EPL.2008, Reuter.NJP.2009}.  Using the
parameters in Table~\ref{table:run_parameters}, the value of the
magnetic field contour in Figure~\ref{fig:fig5} is about 2 Gauss,
which should be measurable in the proposed experiment using
Hall-effect sensors.

The experimenter's ability to vary the fluid's magnetic Prandtl number
can be used to explore the onset of turbulence and how that turbulence
affects laminar dynamo action.  The energies of two other simulations
which use the boundary condition presented in Figure~\ref{fig:fig3},
with $Rm=300$ but with $Pm=2.0, 0.5$ ($Re = 150, 600$), are plotted in
Figure~\ref{fig:fig4}.  For $Pm=2.0$, the boundary condition results
in a velocity field that is magnetically unstable.  This flow has a
higher critical magnetic Reynolds number than the $Pm=1.0$ case,
$Rm_{\rm crit} \simeq 274$, and consequently has a lower magnetic
field growth rate, since as expected the growth rates are proportional
to $Rm - Rm_{\rm crit}$.  Interestingly, the magnetic field energy
saturates at essentially the same value as the $Pm = 1.0$ case,
indicating that the saturated magnetic field does not follow the $B^2
\sim (\rho \nu / \sigma a^2) (Rm - Rm_{\rm crit})$ scaling, as might
otherwise be expected~\citep{Petrelis.EPJB.2001}.  The $Pm=0.5$ case
does not magnetically self-excite, with the energy used to initialize
the magnetic field modes quickly dissipating away.  In this case the
flow contains several non-axisymmetric, though stationary, components
which interfere with the growth of the magnetic field.  This
underlying symmetry-breaking hydrodynamic instability is a first step
towards developed turbulence, and as such permits the study of how
non-axisymmetric contributions change the onset conditions for dynamo
action.

\subsection{Turbulent dynamo simulations}
\label{section:turbulent}

The question of the role of turbulence in the development and
maintenance, or hindrance and destruction, of astrophysical magnetic
fields is of ongoing importance.  This is especially true in light of
the recent turbulent-dynamo results from the VKS2
experiment~\citep{Monchaux.PRL.2007} and the suggestion that coherent
turbulence may be responsible for that
dynamo~\citep{Laguerre.PRL.2008}.  Clearly the proposed experiment
will be most relevant to the study of astrophysical dynamos if it is
able to generate self-excited magnetic fields under turbulent
conditions.

\begin{figure}
\plotone{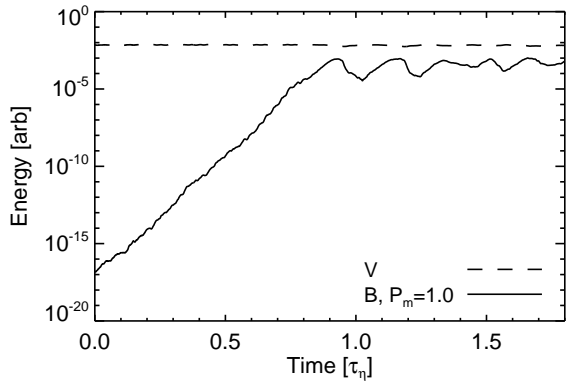}
\caption{Kinetic and magnetic energies versus time for the system
  forced by the boundary condition given in Figure~\ref{fig:fig3},
  with $Rm = 2000$ and $Pm = 1.0$.}
\label{fig:fig6}
\end{figure}

To examine this question, simulations were performed at the higher end
of the experiment's expected range of magnetic Reynolds number, $Rm =
2000$, with $Pm = 1.0$ ($Re = 2000$), using the toroidal outer
boundary condition presented in Figure~\ref{fig:fig3}.  The kinetic
and magnetic energies of this simulation versus time are presented in
Figure~\ref{fig:fig6}.  The magnetic energy grows exponentially in
time, saturating an order of magnitude below the kinetic energy.
Using the parameters in Table~\ref{table:run_parameters}, the magnetic
field in saturation peaks as high as 12 Gauss.  Since the magnetic
field varies rapidly in time it should be easily measurable using
B-dot coils.

\begin{figure}
\plottwo{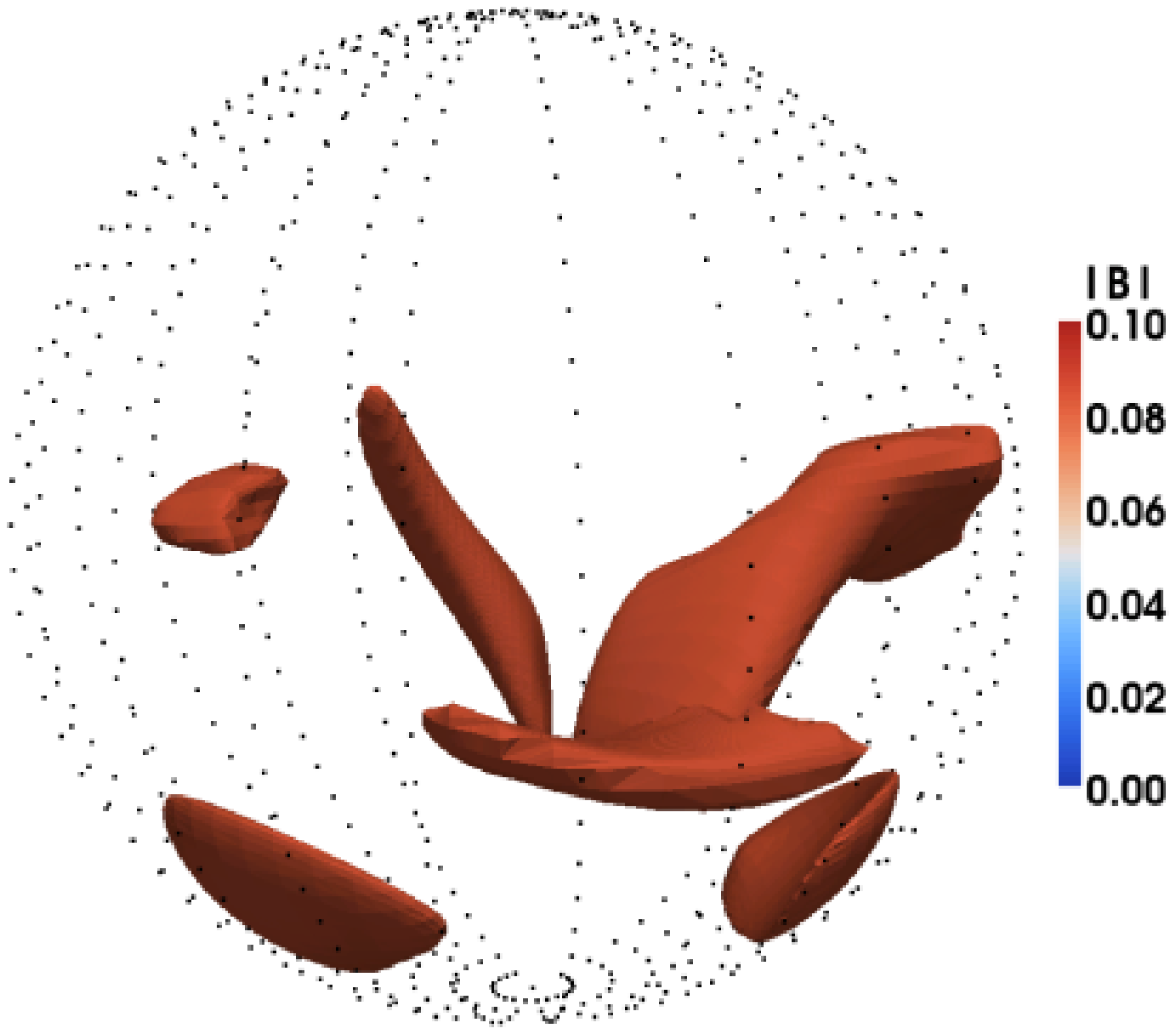}{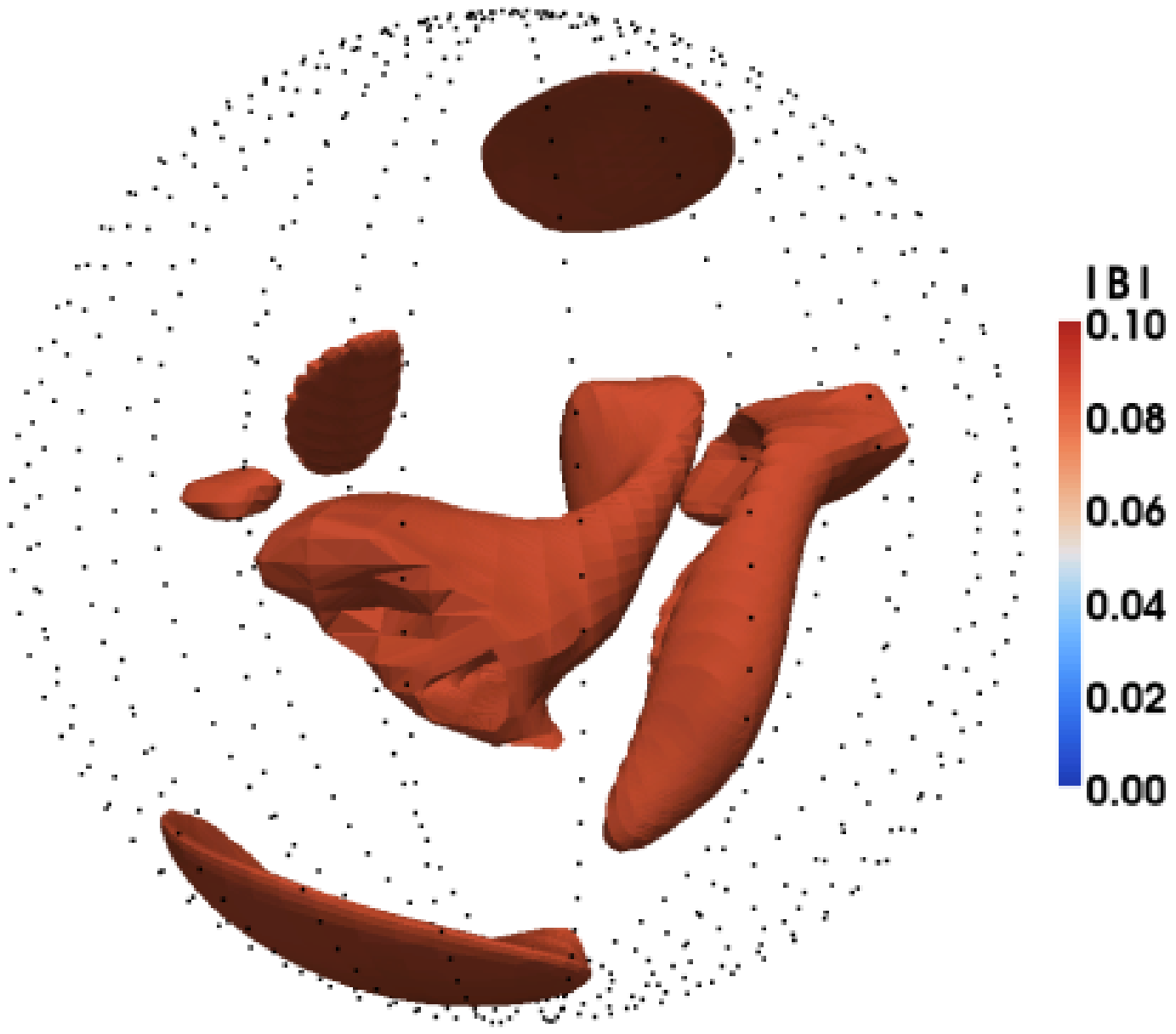}\\
\plottwo{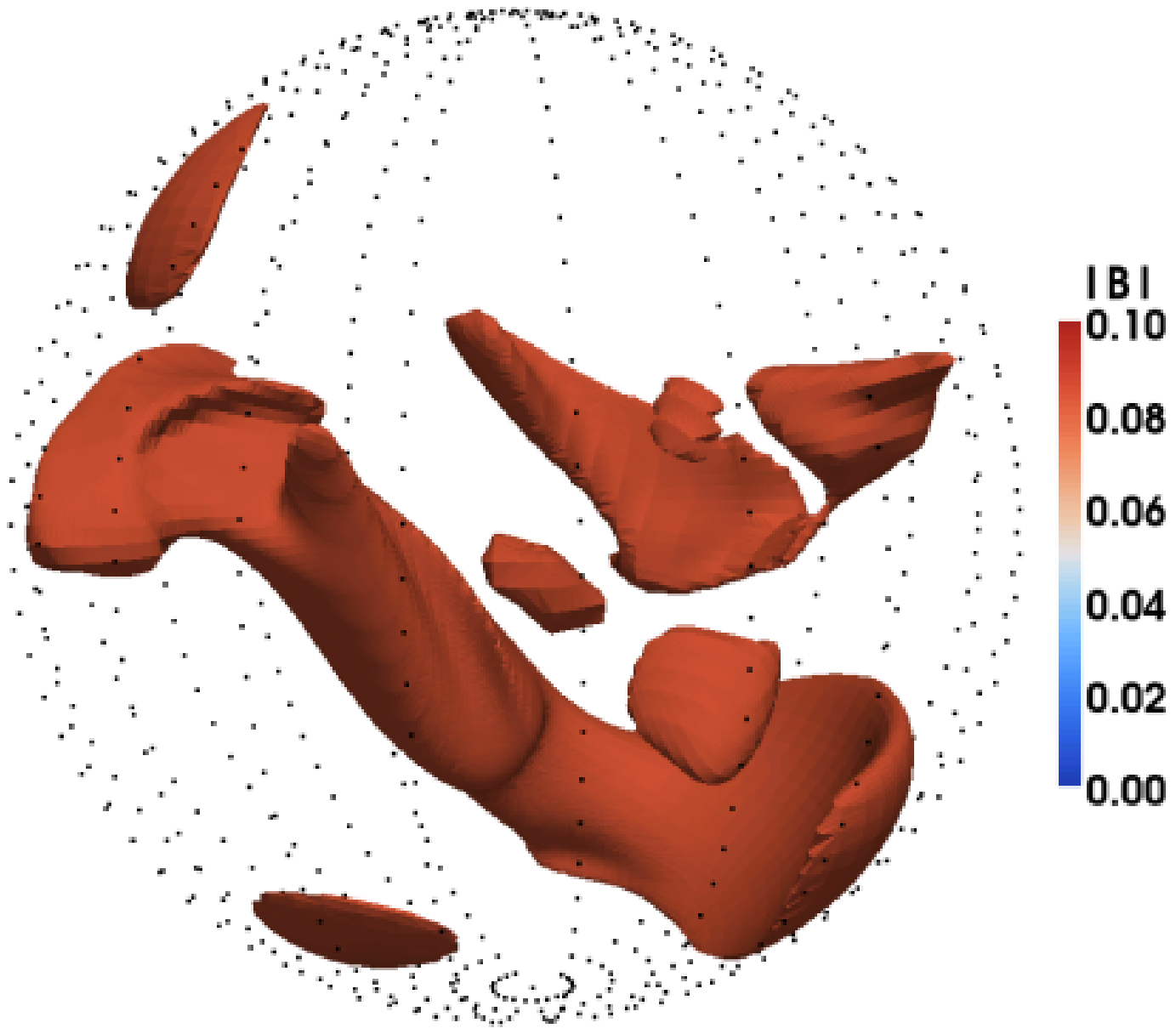}{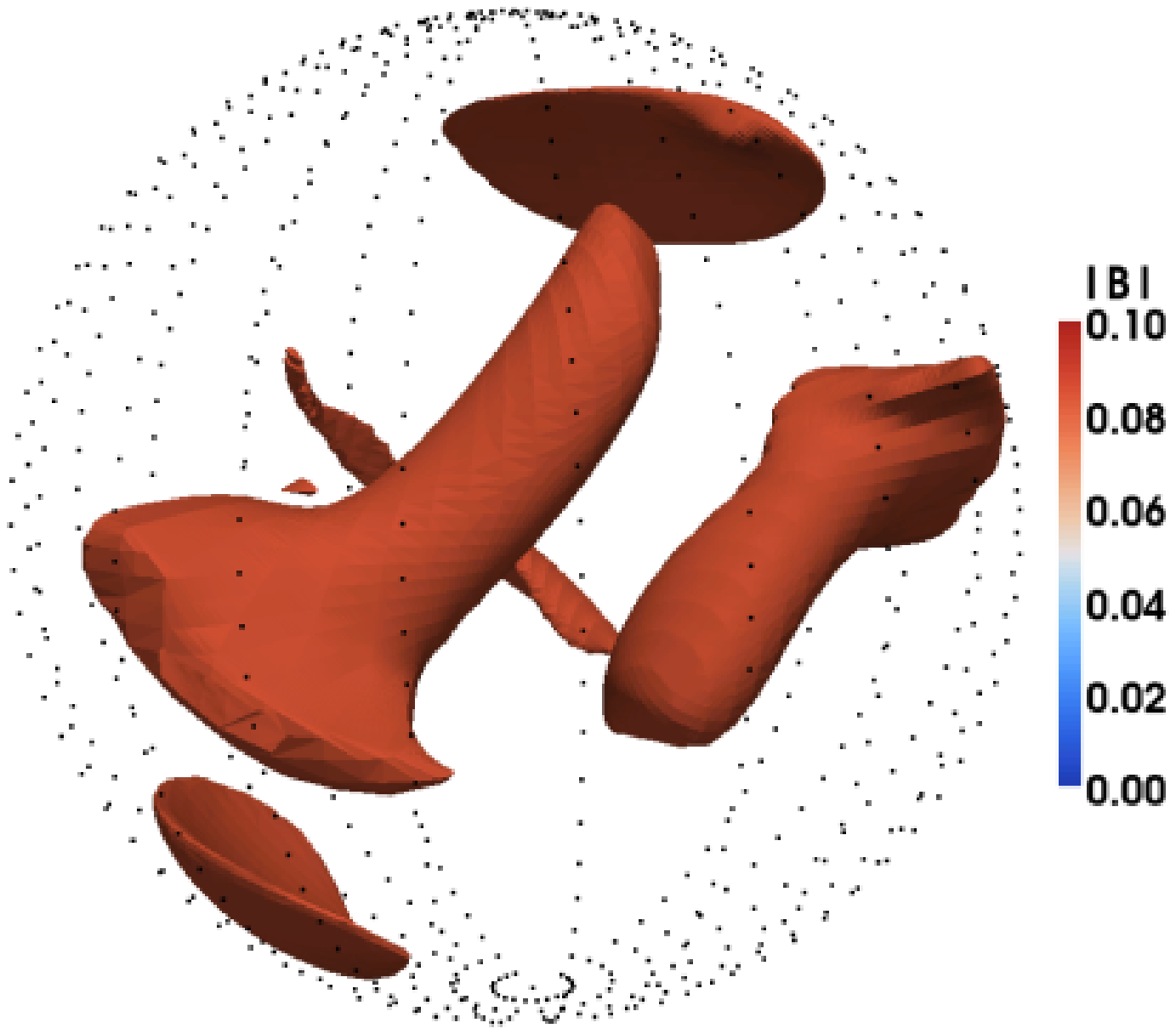}
\caption{Contours of constant magnetic field magnitude for four
  different times during the saturation of the $Rm=2000$, $Pm=1.0$
  turbulent dynamo.  The figures are $\Delta t \simeq 0.2\tau_\eta$
  apart in time.}
\label{fig:fig7}
\end{figure}

\begin{figure}
\plotone{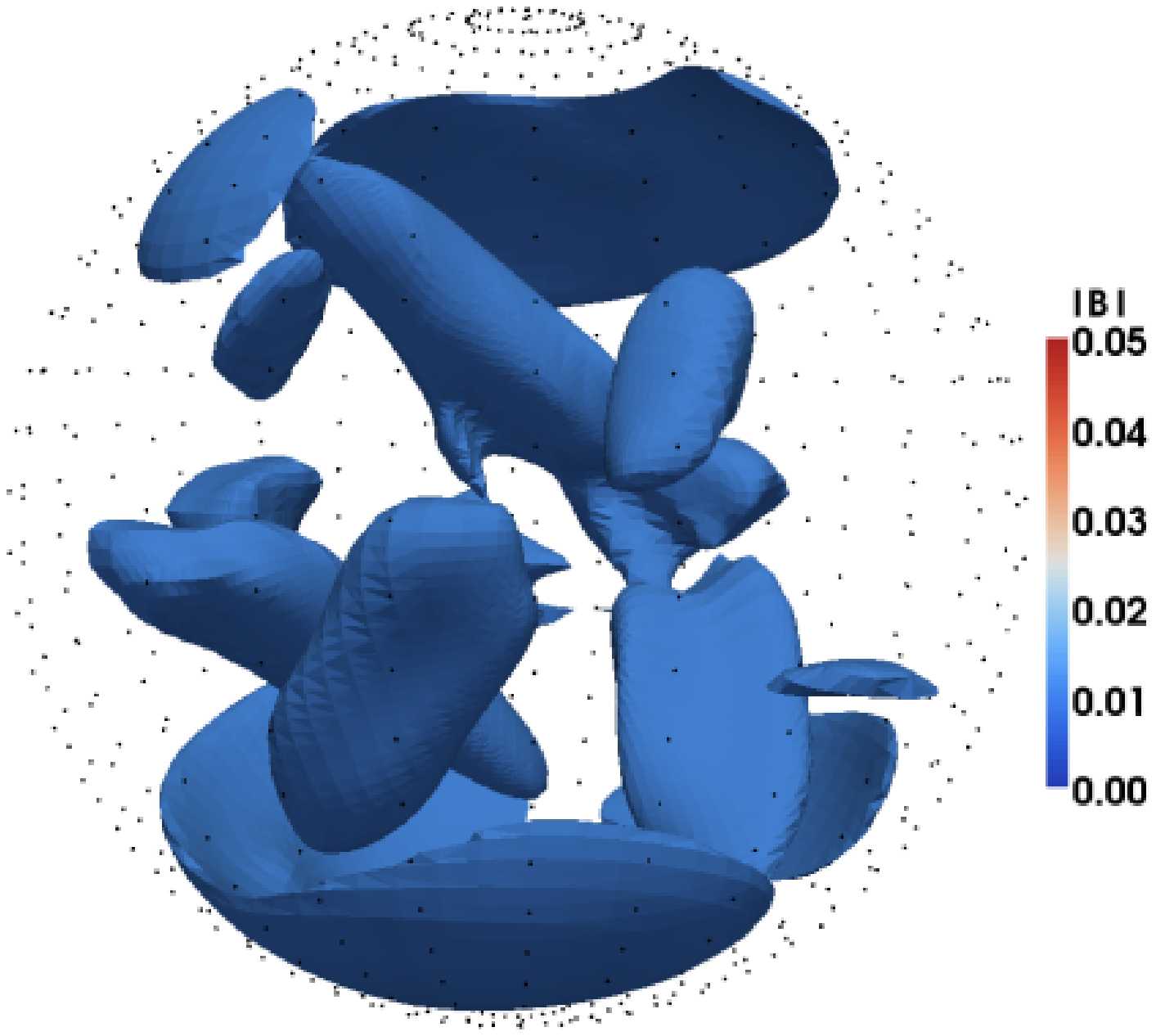}
\caption{Contours of constant magnetic field magnitude for the
  $Rm=2000$, $Pm=1.0$ turbulent dynamo's mean saturated magnetic
  field.  The scale has been reduced from Figure~\ref{fig:fig7} for
  clarity.}
\label{fig:fig8}
\end{figure}

Both the velocity and magnetic fields fluctuate wildly throughout the
simulation.  Snapshots of contours of $|\mathbf{B}|$ during saturation
are presented in Figure~\ref{fig:fig7}.  The transient nature of the
magnetic field, and the lack of obvious large-scale structure at these
contour levels, as compared to the large-scale dynamo
(Figure~\ref{fig:fig5}), is clear.  Nonetheless, the mean saturated
magnetic field is non-zero; its energy is about 4\% of the mean energy
of the magnetic field in saturation, when averaged over 2.9 resistive
times.  Contours of the mean saturated magnetic field are presented in
Figure~\ref{fig:fig8}.  The field is somewhat complex, dominated by
$m=0,1,2$ modes, and is weak at the center of sphere.  The magnetic
field at the sphere's surface peaks at the equator, though it is
composed of $m=1$ and $m=2$ components, in contrast to the large-scale
dynamo, which only consists of $m=1$ modes.

The growth rate of the magnetic field is much larger in this case than
in the laminar cases examined above.  This may be true because the
system is much higher above the critical magnetic Reynolds number for
this set of parameters, as compared to the laminar case, or because a
different dynamo mechanism is at work, perhaps amplification at small
scales.  If the growth rate of the magnetic field were being dominated
by turbulent action at small scales, following
\citet{Batchelor.PRSLA.1950}, we would expect the magnetic energy to
grow exponentially in time as $\left|\mathbf{B}\right|^2 \sim {\rm
  exp}((\epsilon Re)^{1/2}t)$, where $\epsilon$ is the power
dissipation per unit mass.  For this simulation $\epsilon \simeq 16$,
so for the small-scale-dynamo mechanism we would expect the growth
rate to be $\sim 180$, not the $\sim 40$ which is actually measured.
The critical magnetic Reynolds number for this simulation has not yet
been found, though we have determined that it is somewhere in the
range $Rm < 700$.  Based on these observations, it seems unlikely that
the action of small-scale eddies is the dominant contribution to the
growth of magnetic energy, but rather since $Rm=2000$ the rapid growth
rate is explained by the high magnetic Reynolds number.

To gain insight into the length scales important to this dynamo, we
examine the average angle-integrated energy spectra of the velocity
and magnetic fields, $E(k)$ and $M(k)$ respectively, where $M(k) =
k^2\int \langle |\mathbf{B}(\mathbf{k})|^2 \rangle d\Omega_k$, during
the growth phase of the simulation, normalized to their respective
total energies.  These are plotted in Figure~\ref{fig:fig9} (the
details of how $\mathbf{B}(\mathbf{k})$ is calculated are given in
Appendix~\ref{app:transform}).  The spectra have a number of
interesting features.  First, the velocity field spectrum is very
noisy at higher values of $k$.  This phenomenon is likely caused by
the transform to $k$-space, which involves a radial integration across
the edge of the sphere, where the velocity field quickly goes to zero
(except for the toroidal boundary condition).  Since the magnetic
field does not go to zero at the sphere's edge it does not display
this effect.  The velocity field spectrum peaks at relatively low
wavenumber, as is expected since the energy in the simulation is being
injected at the largest scales.  The magnetic energy, in contrast,
peaks at higher wavenumber and drops off rapidly.  The general
understanding of small-scale dynamos, at least for $Pm \sim 1$ and $Pm
\gg 1$~\citep{Schekochihin.APJ.2004} is that the magnetic spectrum
during the kinematic phase should peak at $k_\eta$, which for $Pm =
1.0$ simulations should be the same as $k_\nu$, the knee in the
velocity field spectrum.  This is not the case in this situation,
again suggesting that this is not a small-scale dynamo.  Of course it
is worth noting that the current understanding of the small-scale
dynamo exists in the context of a number of assumptions, most notably
those of homogeneous and isotropic turbulence and the lack of a mean
flow; none of these assumptions apply in this situation.
\begin{figure}
  \plotone{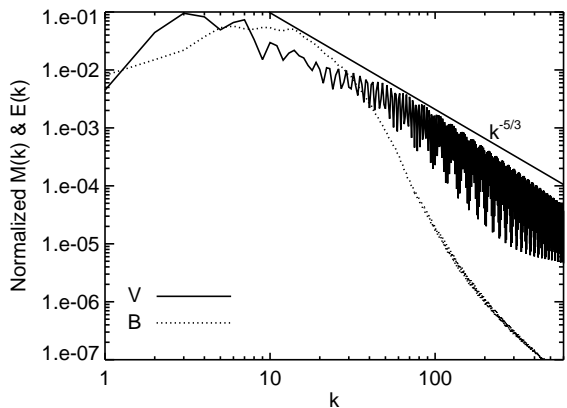}
  \caption{Time-averaged, angle-integrated, kinematic magnetic and
    kinetic energy spectra, $M(k)$ and $E(k)$ respectively, for the
    $Rm=2000$, $Pm=1.0$ dynamo, normalized to their respect energies.}
  \label{fig:fig9}
\end{figure}

Though it has not been ruled out definitely, based on the growth rate
of the magnetic energy, the non-zero mean saturated magnetic field,
and the spectra of the kinematic magnetic field, we believe that the
dynamo presented here is not a small-scale dynamo, but rather is an
example of a turbulent large-scale dynamo.  It is possible that this
dynamo is related to some of the previously studied turbulent dynamos
which possess a mean flow and similar magnetic
spectra~\citep{Ponty.PRL.2005, Mininni.PP.2006}, or the
recently-discovered 'shear dynamo'~\citep{Yousef.PRL.2008}, though
none of these studies were done in a finite domain.

\section{Discussion and conclusions}

The simulations of the proposed experiment presented here invoke a
number of assumptions regarding the forcing at the plasma's outer
edge.  Though we believe that these simulations capture the important
physics of the experiment, the assumption that the forcing can be
modeled using a continuous no-slip boundary condition is debatable.
Clearly local effects caused by the discrete magnetic and electric
fields are being ignored with this treatment.  Future simulations of
the experiment will endeavor to more accurately model the forcing by
using the NIMROD code~\citep{Sovinec.JCP.2004} to add two-fluid,
compressible, and collisionless physics, including anisotropic and
long-mean-free-path effects.  Such additions may be especially
important at lower values of magnetic Prandtl number, as one might
expect the development of Stewartson layers
\citep{Stewartson.JFM.1957} near the discrete areas of forcing,
similar to effects found in cylindrical MRI experiments with
differentially rotating upper and lower rings~\citep{Burin.EF.2006}.
One might also expect the plasma to transport cusp magnetic field into
the volume of the sphere, a phenomena that has been observed with
flowing liquid sodium~\citep{Volk.PRL.2006}.  This effect could be
important for rotating convection experiments, especially for the
study of the physics of the tachocline~\citep{Miesch.LRSP.2005}.

Since it has long been known that many axisymmetric flows exist which
magnetically self-excite \citep{Roberts.Paris.1971,
  Gubbins.PRTSLA.1973, Dudley.PRSLA.1989} it is expected that there
are a multitude of boundary conditions which could be programmed into
the experiment which would result in laminar dynamo action, allowing
the study of the different means by which large-scale dynamos occur.
Other boundary conditions which might be of interest include
solar-type boundary conditions, where the equator spins much faster
than the poles, and gas-giant-type boundary conditions, where the
surface contains many prograde and retrograde jets.  Density
stratification and convection in rapidly rotating plasmas are also
possible for this experiment, and should be studied.

The laminar dynamo presented in this study is large scale, as must be
the case when there is no energy in the velocity field at small
scales.  As the fluid Reynolds number is increased the simulations
become turbulent.  As described in Section~\ref{section:turbulent},
the current understanding of the turbulent dynamo presented here is
that it is not a small-scale dynamo, but rather a very turbulent
large-scale dynamo.  Because the outer toroidal boundary condition can
be controlled in time, it is conceivable that a turbulent small-scale
dynamo could be generated in this experiment without the presence of a
mean flow.  This is a topic which requires further study.  Another
question of considerable interest is: at what point does a large-scale
dynamo become so turbulent that it ceases to have appreciable energy
at the large scales and becomes small-scale?  Does such a transition
exist, or does the presence of a mean flow prevent this from
happening?  The proposed experiment, with its ability to reach high
values of $Rm$ and wide range of kinetic Reynolds number, would be
uniquely positioned to study these questions.

It is easily possible to reach $Pm \gg 1$ under laminar conditions, as
been shown in many numerical
simulations~\citep{Schekochihin.APJ.2004}.  Another goal of the
proposed experiment should be to attempt to study the $Pm \gg 1$
regime under fully turbulent conditions.  This will be challenging, at
least for the boundary condition examined here.  The forcing strategy
of the proposed experiment does not fill the volume of the fluid, but
only affects the boundary, and thus a high value of $Re$ is needed to
make the fluid turbulent.  (A simulation was performed with $Rm=2000$
and $Re=1000$.  This was laminar until saturation of the dynamo was
achieved.)  This need for a high value of fluid Reynolds number puts
an upper limit on the experimental value of $Pm$ that can be reached
under turbulent conditions.  Those researchers that perform
simulations of this physical regime suffer from a similar problem; for
the simulators the problem is a lack of the resolution needed to reach
high $Pm$ at high $Rm$, for the proposed experiment it is one of a
technical upper limit on the value of magnetic Reynolds number that is
experimentally accessible.

In summary, we have presented a concept for a high-$\beta$ plasma
experiment which by its nature would be ideal for studying the MHD
dynamo and a variety of astrophysical fluid dynamics phenomena, many
of which have never been studied experimentally.  The experiment's
plasma confinement is based on a ring-cusp strategy, and momentum is
injected into the plasma via differential toroidal $\mathbf{E}
\boldsymbol{\times} \mathbf{B}$ forcing at the plasma's outer edge.
With the advances in technology of the last few decades this
experiment should be able to reach parameter regimes inaccessible to
previous multidipole experiments.  We have demonstrated through
numerical simulations that such a control scheme is sufficient for
generating velocity fields which are capable of both laminar and
turbulent dynamo action.  The combination of velocity-field
tunability, high plasma $\beta$, and wide range of kinetic and
magnetic Reynolds numbers would make this experiment a viable choice
for exploring a multitude of astrophysical phenomena.

\acknowledgments

EJS thanks Dr.\ M. Nornberg for helpful conversations.  This work was
made possible by the facilities of the Shared Hierarchical Academic
Research Computing Network (SHARCNET:www.sharcnet.ca).

\appendix

\section{Outer Boundary Values}
\label{app:boundary_values}

The toroidal velocity field radial profile boundary values which are
used to generate the results presented in this paper, and which are
used to generate Figure~\ref{fig:fig3}, can be found in
Table~\ref{table:outer_boundary_values}.
\begin{deluxetable}{ccccc} 
\tablecolumns{5} 
\tablewidth{0pc} 
\tablecaption{Toroidal Boundary Values} 
\tablehead{ 
\colhead{$\ell$} & \colhead{Fig.~\ref{fig:fig2}(a)} & 
\colhead{Fig.~\ref{fig:fig2}(b)} & 
\colhead{Fig.~\ref{fig:fig2}(c)} & 
\colhead{Fig.~\ref{fig:fig2}(d)}}
\startdata 
1      & 0.5773  & 0.2273  & -0.0667  &  0.0000 \\
2      & 0.0000  & 0.0000  &  0.0000  & -0.1002 \\
3      & 0.0000  & 0.0541  & -0.1234  &  0.0000 \\
4      & 0.0000  & 0.0000  &  0.0000  & -0.0788 \\
5      & 0.0000  & 0.0239  & -0.0130  &  0.0000 \\
6      & 0.0000  & 0.0000  &  0.0000  & -0.0110 \\
7      & 0.0000  & 0.0124  &  0.0279  &  0.0000 \\
8      & 0.0000  & 0.0000  &  0.0000  &  0.0117 \\
9      & 0.0000  & 0.0067  &  0.0000  &  0.0000 \\
11     & 0.0000  & 0.0036  &  0.0000  &  0.0000 \\
13     & 0.0000  & 0.0018  &  0.0000  &  0.0000 \\
15     & 0.0000  & 0.0008  &  0.0000  &  0.0000 \\
17     & 0.0000  & 0.0002  &  0.0000  &  0.0000 \\
\enddata 
\label{table:outer_boundary_values}
\end{deluxetable} 
It should be noted that the simulation code uses the following
normalization for axisymmetric spherical harmonics:
\begin{equation}
Y_\ell^0(\theta,\phi) = \sqrt{(2\ell+1)}P_\ell^0(\cos\theta).
\end{equation}
This normalization affects the radial profile boundary values.

\section{Spatial Transforms}
\label{app:transform}

The simulation code evolves the velocity and magnetic fields in a
spherical harmonic basis.  To calculate $M(k)$ and $E(k)$ the fields
must be transformed into spherical $k$ space.  Rather than convert
from the spherical harmonic basis to physical space, and then
transform to $k$ space, the fields are transformed directly from the
spherical harmonic basis to spherical $k$ space.  This Appendix
outlines how this is accomplished.

The native form of the fields is in terms of radial profiles projected
onto a spherical harmonic basis, assuming that the fields are
divergence free:
\begin{equation}
\mathbf{B} = \sum_\alpha \boldsymbol{\nabla} \boldsymbol{\times}
\boldsymbol{\nabla} \boldsymbol{\times}
\left[S_\alpha(r)Y_\alpha(\theta,\phi)\mathbf{r}\right] +
\boldsymbol{\nabla} \boldsymbol{\times}
\left[T_\alpha(r)Y_\alpha(\theta,\phi)\mathbf{r}\right],
\end{equation}
where we follow the convention of~\citet{Moffatt} and use a full
$\mathbf{r}$ vector, as opposed to the convention
of~\citet{Bullard.PTRSLA.1954} who used the $\mathbf{\hat{r}}$ unit
vector.  The summation over $\alpha$ is over all valid spherical
harmonic combinations, $\ell$ and $m$, starting at $\ell=1$, truncated
at some $\ell_{\rm max}$ and $m_{\rm max}$.  The radial profiles
$S_\alpha(r)$ and $T_\alpha(r)$ are the profiles which characterize
the field.  Thus defined, the components of the field take the form
\begin{align}\label{eqn:br}
B_r(r,\theta,\phi) & = \sum_\alpha \frac{\ell_\alpha(\ell_\alpha+1)
  S_\alpha(r)}{r} Y_\alpha(\theta,\phi),\\\label{eqn:btheta}
B_\theta(r,\theta,\phi) & = \sum_\alpha
\left[\frac{1}{r}\frac{\partial (rS_\alpha(r))}{\partial
    r}\frac{\partial Y_\alpha}{\partial\theta} +
  \frac{T_\alpha(r)}{\sin\theta}\frac{\partial Y_\alpha}{\partial\phi}
  \right], \\\label{eqn:bphi} B_\phi(r,\theta,\phi) & = \sum_\alpha
\left[ -T_\alpha(r)\frac{\partial Y_\alpha}{\partial\theta} +
  \frac{1}{r\sin\theta}\frac{\partial (rS_\alpha(r))}{\partial
    r}\frac{\partial Y_\alpha}{\partial\phi} \right].
\end{align}
We now desire
\begin{equation}\label{eqn:transform}
\mathbf{B}(\mathbf{k}) = \frac{1}{(2\pi)^{3/2}}\int
e^{i\mathbf{k}\cdot\mathbf{x}} \mathbf{B}(\mathbf{x}) d\mathbf{x} =
\frac{1}{(2\pi)^{3/2}}\int e^{ikr\cos\gamma} \mathbf{B}(\mathbf{x})
r^2dr \sin\theta d\theta d\phi,
\end{equation}
where $\cos\gamma = \sin\theta_k\sin\theta\cos\left(\phi-\phi_k\right)
+ \cos\theta_k\cos\theta$.  Calculation of this integral requires
several identities.  We will use the Rayleigh equation,
\begin{equation}\label{eqn:bessel}
e^{ikr\cos\gamma} = \sum_{n=0}^\infty
i^n\left(2n+1\right)j_n(kr)P_n(\cos\gamma),
\end{equation}
where $j_n(kr)$ is the spherical Bessel function of the first kind,
and $P_n(\cos\gamma)$ is the Legendre polynomial.  This expansion over
$n$ is truncated at $\ell_{\rm max}$, which gives satisfactory
convergence for the cases considered here.  We also need the addition
theorem for spherical harmonics,
\begin{equation}\label{eqn:addition}
P_n(\cos\gamma) = \sum_{m=0}^{m=n} \frac{4\pi}{2n+1}
Y_n^{m\ast}(\theta,\phi) Y_n^m(\theta_k,\phi_k).
\end{equation}
We first consider the radial component of this transform.  Combining
equations~\ref{eqn:br}, \ref{eqn:transform}, \ref{eqn:bessel} and
\ref{eqn:addition} gives
\begin{equation}
B_k(k,\theta_k,\phi_k) = \frac{2}{\sqrt{2\pi}}\sum_{n=1}^{\ell_{\rm
    max}} \sum_{m=0}^{m=n} i^nn\left(n+1\right) Y_n^m(\theta_k,\phi_k)
\int_0^\infty j_n(kr) S_n^m(r) r dr,
\end{equation}
where orthonormality between the spherical harmonics has been assumed.
The $\theta_k$ and $\phi_k$ components of the transform are a little
more complicated.  Let us consider the first term on the right hand
side of equation~\ref{eqn:btheta}.  If the spherical harmonics are
defined as $Y_\alpha(\theta,\phi) = N_\alpha
P_{\ell_\alpha}^{m_\alpha}(\cos\theta) e^{im_\alpha \phi}$, where
$N_\alpha$ is the normalization constant, then the contribution from
the first term is given by
\begin{align}\nonumber
B_{\theta_k}(k,\theta_k,\phi_k)_{\rm 1^{\rm st}} =
\frac{4\pi}{\sqrt{2\pi}} \sum_{n=0}^{\ell_{\rm max}}
\sum_{\ell_\alpha=1}^{\ell_{\rm max}} \sum_{m=0}^{m=n} & N_n N_\alpha
i^n Y_n^m(\theta_k,\phi_k) \int_0^\infty j_n(kr) \frac{\partial
  (rS_{\ell_\alpha}^m)}{\partial r} rdr \times \\ & \int_0^\pi
P_n^m(\cos\theta) \frac{\partial
  P_{\ell_\alpha}^m(\cos\theta)}{\partial\theta} \sin\theta d\theta.
\end{align}
The second term is similar:
\begin{align}\nonumber
B_{\theta_k}(k,\theta_k,\phi_k)_{\rm 2^{\rm nd}} =
\frac{4\pi}{\sqrt{2\pi}} \sum_{n=0}^{\ell_{\rm max}}
\sum_{\ell_\alpha=1}^{\ell_{\rm max}} \sum_{m=0}^{m=n} & N_n N_\alpha
i^{n+1} m Y_n^m(\theta_k,\phi_k) \int_0^\infty j_n(kr)
T_{\ell_\alpha}^m(r) r^2 dr \times \\ & \int_0^\pi P_n^m(\cos\theta)
P_{\ell_\alpha}^m(\cos\theta) d\theta.
\end{align}
The terms needed to calculate $B_{\phi_k}$ are similar, the only
differences being the quantities in the radial integrals.

It should be observed that the radial integral is evaluated all the
way to infinity.  For transforms of the velocity field, this integral
is only evaluated up to $r=1$, since that is where the radial profiles
go to zero, with the exception of the toroidal radial profiles which
have non-zero boundary conditions.  (This truncation of the radial
profile at $r=1$ is what gives the velocity field spectra their spiky
nature at high $k$.  See Figure~\ref{fig:fig9} for an example.)  The
radial integrals over $S_\alpha(r)$ are evaluated all the way to
infinity, since the poloidal radial profiles are non-zero outside the
sphere.  Since the region $r > 1$ is current-free, the radial profile
is matched to the vacuum solution for the magnetic field.  The
external part of the radial integrals then become
\begin{align}
\int_1^\infty j_n(kr) S_n^m(r) r dr & = S_n^m(1) \int_1^\infty j_n(kr)
r^{-n} dr,\\ \int_1^\infty j_n(kr) \frac{\partial
  (rS_{\ell_\alpha}^m(r))}{\partial r} r dr & = -\ell_\alpha
S_{\ell_\alpha}^m(1) \int_1^\infty j_n(kr) r^{-\ell_\alpha} dr,
\end{align}
the former integral being a special case of the latter.  These
integrals don't change, and so have been tabulated for repeated use.
The toroidal part of the magnetic field, $T_\alpha(r)$, like the
velocity field radial profiles, goes to zero at $r=1$, since we are
assuming an insulating outer boundary.  Thus, the radial integrals
over $T_\alpha(r)$ are only evaluated up to $r=1$.

\end{document}